\documentclass[journal,letterpaper]{IEEEtran}

\usepackage{graphicx,cite,epsfig,amssymb,amsmath,amsfonts,multicol,subfigure,mathtools,bm,mathrsfs,setspace}
\usepackage{multirow}

\newcommand{\tabincell}[2]{\begin{tabular}{@{}#1@{}}#2\end{tabular}}

\begin{document}
\title{\huge A Signal Separation Method for Exceeding the Channel Capacity of BPSK Input}
\author{
	
	Bingli~Jiao,
Dongsheng Zheng and
Yuli Yang
\thanks{B. Jiao ({\em corresponding author}) is with the Department of Electronics and Peking University-Princeton University Joint Laboratory of Advanced Communications Research, Peking University, Beijing 100871, China (email: jiaobl@pku.edu.cn).}
\thanks{D. Zheng is with the Department of Electronics and Peking University-Princeton University Joint Laboratory of Advanced Communications Research, Peking University, Beijing 100871, China (email: zhengds@pku.edu.cn).}
\thanks{Y. Yang is with the Department of Electronic and Electrical Engineering, University of Chester, Chester CH2 4NU, U.K. (e-mail: y.yang@chester.ac.uk).}

}

\maketitle

\begin{abstract}

The field of the coded modulations witnessed its golden era as the simulated achievable bit rates (ABRs) were getting close to the channel capacities of the finite alphabet inputs.  However, when working at a single channel, the previous researches are mostly restricted in scope of single input and single output.  In contrast, for achieving higher spectral efficiency, the present authors propose a method that builds upon the parallel transmission of two signals that are separated at the receiver.   The signal separation is enabled by, so called, the double mapping modulation working through Hamming- to Euclidean space.  Consequently, the ABR summation of the two signals is found higher the channel capacity of BPSK input. Both the theoretical- and simulation result confirm this approach.

\end{abstract}

\begin{IEEEkeywords}
achievable bit rate, coded modulation, channel capacity. 
\end{IEEEkeywords}

\IEEEpeerreviewmaketitle

\section{Introduction}
In the past years, one of the practical researches on capacity issues was on the design of the coded modulation working over the additive white Gaussian noise (AWGN) channel
\begin{eqnarray}
\begin{array}{l}\label{equ-ch}
y = x + n,
\end{array}
\end{eqnarray}
where $y$ and $x$ are the output- and input signal of the channel, and $n$ is the received AWGN component from a normally distributed ensemble of power $\sigma_N^2$ denoted by $n \sim \mathcal{N}(0,\sigma_N^2)$.  

At the finite alphabet input, the Shannon channel capacities are calculated by the mutual information \cite{Shannon1948}  
\begin{equation}
\begin{array}{l}\label{equ1}
\rm{I}(X;Y) =\rm{H}(Y) - \rm{H}(N)
\end{array}
\end{equation}
where $\rm{I}(X;Y)$ is the mutual information, ${\rm{H}}(Y)$ is the entropy of the received signal,  ${\rm{H}}(N) = {\log _2} (\sqrt{2 \pi e \sigma_N^2})$ is that of the AWGN \cite{Verdu2007}.  The numerical results of \eqref{equ1} are shown in Fig. 1.

In the practical study, one can target a value of the mutual information by designing the coded modulation scheme  \cite{Ungerboeck1982,Poor2011} and simulate the bit error rate (BER) performance, where the BER at $10^{-6}$ or $10^{-8}$ is regarded as an approximation of  ``arbitrarily small error probability" whereat bit-energy to noise ratio is used to the comparison at horizontal axis with the mutual information.  

More accurate estimation of approaching the channel capacity requires the theoretical analysis based on asymptotic investigations. A good example can be found from the research of BPSK plus low-density parity-check (LDPC) code  that approaches the channel capacity at 0.5 bits/s/Hz at a gap of 0.0045dB \cite{Chung2001}.      

\begin{figure}[htb]
	\centering
	\includegraphics[width=0.4\textwidth]{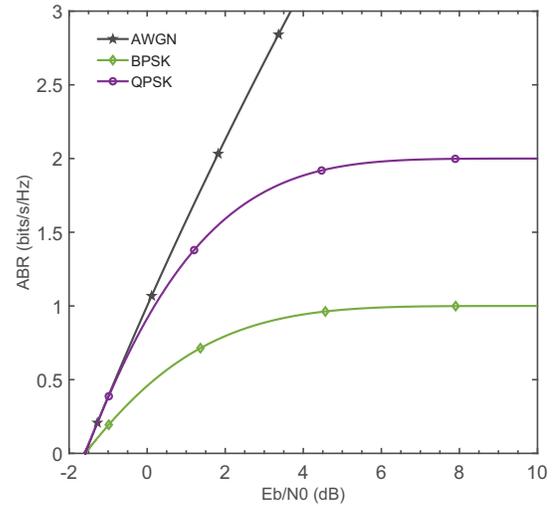}
	\caption{Mutual Information of BPSK and QPSK}
	\label{Fig3}
\end{figure}

In contrast to the above well-developed approach, the present authors propose a signal separation method that divides \eqref{equ-ch} into two 
\begin{eqnarray}
\begin{array}{l}\label{s-1}
y_1 = x_1 + n,
\end{array}
\end{eqnarray}
and 
\begin{eqnarray}
\begin{array}{l}\label{s-2}
y_2 = x_2 + n,
\end{array}
\end{eqnarray}
for increasing the spectral efficiency further, where $x_1$ and $x_2$ are the separated signals and $y_1$ and $y_2$ are the corresponding received signals \cite{jiao}.

To target ABR of the conventional BPSK, we use BPSK modulation to $x_1$ and create a new modulation, referred to as the double mapping modulation, to work with $x_2$ for making it separable from $x_1$.  By such, signal $x_2$ can transmit additional information bits without consuming any more power in comparison with the conventional BPSK.  Consequently, one can find that $x_2$ brings a direct ABR gain and contributes to bit-energy to noise ratio as well.  The simulation results substantiate the above theoretical envisions.

Throughout the present paper, we use the capital letter to express a vector in Hamming space and the small letter to indicate its component, e.g.,  $A = \{a_1,a_2, ...., a_M\}$, where $A$ represents the vector and $a_i$ the $ith$ component, and use bold face letter to express the signals in Euclidean space with two dimensional complex plan.  In the derivations, we use $\hat{y}$ to express the estimate of $y$ at the receive through out the paper.

\section{Communication Scheme}
Let us consider two independent binary source bit sequences, which are expressed
in vector form of $C^{(i)} = \{c^{(i)}_1,c^{(i)}_2,..,c^{(i)}_{k_i},...,c^{(i)}_{K_i}\}$, where  $c^{(i)}_{k_i}$ is the $kth$ information bit (info-bit) of $C^{(i)}$ and $K_i$ is the length of each source subsequence, and $i=1,2$ indicates the two sources, respectively. 

The two source bit sequences are encoded to  $V^{(i)} = \{v^{(i)}_1,v^{(i)}_2,..,v^{(i)}_m...,v^{(i)}_{N}\}$ by using two different channel code matrices
\begin{eqnarray}
\begin{array}{l}\label{equ-code-1}
v^{(i)}_{m} = \sum\limits_{k_i}g^{(i)}_{{m}k_i}c^{(i)}_{k_i}  
\end{array}
\end{eqnarray}  
where $v^{(i)}_m$ is the $mth$ code-bit of $V^{(i)}$ with $v^{(i)}_m \in (1,0)$ and $g^{(i)}_{mk_i}$ is the element of the code matrix $G^{(i)}$. 

\subsection{Signal Modulation}   
For simplification, the series' indices of the code-bits of $V^{(1)}$ and $V^{(2)}$ are omitted in our organization of the two signals in the following derivations.  

First, we define $x_1$ by using the conventional BPSK, in the complex plan, to the modulation of each bit from $V^{(1)}$ as 
\begin{eqnarray}
\begin{array}{l}\label{x_1}
\textbf{x}_1 = (\sqrt{E_{x_s}},j0)   \ \ \ \ \  \ for \ \ v^{(1)}=0  \\
\textbf{x}_1 = (-\sqrt{E_{x_s}}, j0)  \ \ \ \ for \ \ v^{(1)}=1   
\end{array}
\end{eqnarray}  
where $E_s$ is the symbol energy of $\textbf{x}_1$ and $j=\sqrt{-1}$.  

Secondly, for constructing $x_2$, we define a rotation operator $\Gamma_{\beta}=e^{j\beta}$ to rotate a vector for angle $\beta$ in complex plane, e.g., 
\begin{eqnarray}
\begin{array}{l}\label{x_2}
\textbf{z}'= \Gamma_{\beta}\textbf{z} =\textbf{z}e^{j\beta}
\end{array}
\end{eqnarray}
where $\textbf{z}'$ is the vector from the rotation of $\textbf{z}$ by angle $\beta$. To complete the operator's explanations, we note that $\Gamma_0=1$ and $\Gamma_{\beta}\Gamma_{-\beta}=1$ hold. 

The code-bits of $V^{(2)}$ are mapped onto $\beta$ one another as follows:  $v^{(2)}=0$ is mapped onto $\beta=0$ and $v^{(2)} =1$ onto $\beta =\pi/2 $.   The signal $x_2$ is converted to a vector version by $\textbf{x}_2$ in the complex plane by 
\begin{eqnarray}
\begin{array}{l}\label{x_2}
\textbf{x}_2= \Gamma_{\beta}\textbf{x}_1 =\textbf{x}_1e^{j\beta}
\end{array}
\end{eqnarray}
where $\textbf{x}_2$ is the modulated symbol carrying one bit by the rotation angle $\beta$.  We refer to $\textbf{x}_2$ as the out-layer signal and $\textbf{x}_1$ as the inner BPSK signal.

\begin{figure}[htb]
	\centering
	\includegraphics[width=0.26\textwidth]{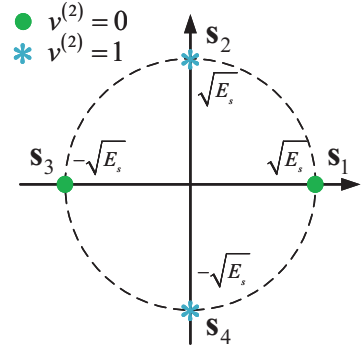}
	\caption{Constellation of the proposed scheme.}
	\label{Fig2}
\end{figure}

\begin{table}[htb]
	\renewcommand{\arraystretch}{1.5}
	\centering
	%\tiny
	\small
	\caption{Signal modulation results.}
	\label{Table1}
	\begin{tabular}{|c|c|c|}
		\hline
		\tabincell{c}{$v^{(2)}$} & \tabincell{c}{$\beta$} & \tabincell{c}{$\textbf{x}_{2}$} \\
		\hline
		\multirow{2}{*}{$0$} & \multirow{2}{*}{$0$} & \tabincell{c}{$\textbf{s}_{1}$} \\
		\cline{3-3}
		& & \tabincell{c}{$\textbf{s}_{3}$} \\
		\hline
		\multirow{2}{*}{$1$} & \multirow{2}{*}{$\pi/2$} & \tabincell{c}{$\textbf{s}_{2}$} \\
		\cline{3-3}
		& &\tabincell{c}{$\textbf{s}_{4}$} \\	
		\hline
		
	\end{tabular}
\end{table}

Since $\textbf{x}_1$ has two possible values, each value of $v^{(2)}$ can be mapped onto two possible points in the complex planes as shown in Fig. 2, where
$\textbf{s}_1$ and $\textbf{s}_3$ are the mapping results of $v^{(2)}=0$,   and $\textbf{s}_2$ and $\textbf{s}_4$ the results of $v^{(2)}=1$.  We refer to the modulation as the double mapping modulation (DMM) because one value of $v^{(2)}$ is mapped onto Euclidean space for the two points as shown in Fig. 2.  The DMM is listed in Table I for the demodulations latter. 

We note that the boldface letter $\textbf{x}_1$ and $\textbf{x}_2$ are used to replace signal $x_1$ and $x_2$ in \eqref{s-1} and \eqref{s-2} and the both signals share the same symbol energy of $E_s$.

Finally, because $\textbf{s}_k$ contains total information bits carried by $\textbf{x}_1$ and $\textbf{x}_2$, the transmitter uses it as the channel input to  
\begin{eqnarray}
\begin{array}{l}\label{equ-code}
\textbf{y}=\textbf{s}_k + \textbf{n} = \Gamma_{\beta}\textbf{x}_1 + \textbf{n}
\end{array}
\end{eqnarray}  
where $\textbf{s}_k$ is the transmit symbol and $\textbf{n}$ is the two dimensional noise.

\subsection{At Receiver}
In the proposed method, the receiver needs to record all signals in a storage for the demodulations of $\textbf{x}_2$ and $\textbf{x}_1$ separately.  Actually, the signals will be used twice: one time for the demodulation of $V^{(2)}$ and the other for $V^{(1)}$.  The procedures of the information recovery are done inversely with respect to that at the transmitter. 

At first, out-layer signal $\textbf{x}_2$ is demodulated by  
\begin{equation}
\begin{array}{l}\label{equ3}
\hat{\textbf{y}} = \textbf{s}_1 \ \ or \ \ \textbf{s}_3  \ \ for \ \ v^{(2)} =0  \\
\hat{\textbf{y}} =\textbf{s}_2 \ \ or \ \ \textbf{s}_4  \ \ for  \ \ v^{(2)}=1  \\
\end{array}
\end{equation}
where $\hat{\textbf{y}}$ is the estimate of $\textbf{y}$ in the demodulation of $v^{(2)}$.  By using a conventional decoding method, one can obtain recovered info-bits denoted as $\hat{C}^{(2)}$. 

Secondly, once $\hat{C}^{(2)}$ has been obtained, the receiver reconstructs all code-bits of $V^{(2)}$ by using \eqref{equ-code-1} and finds each value of $\hat{\beta}$ in table I.  

Then, the demodulation of $\textbf{x}_1$ can be done by    
\begin{eqnarray}
\begin{array}{l}\label{equ-recove}
\textbf{y}_1= \Gamma_{-\hat{\beta}}\textbf{y} =
\Gamma_{-\hat{\beta}}\Gamma_{\beta}\textbf{x}_1 + \textbf{n}'
\end{array} 
\end{eqnarray}
where $\textbf{y}_1$ is the recovered signal for demodulation of the inner BPSK, $\hat{\beta}$ is the estimate of the rotation angle related to the reconstructed bit value $v^{(2)}$.

Then, one can use a decoding method and recover the info-bits carried by $\textbf{x}_1$ denoted as $\hat{C}^{(1)}$.  

We note that the symbol energy $E_s$ is used for the demodulation of $v^{(2)}$ one time and is reused for $v^{(1)}$ another time. While the SNRs with the two demodulations are the same because the rotation operator does not change the amplitude of the signal and the variance of the noise as well.

\section{Beyond the Channel Capacity}
In this section, we show the advantage of this approach for achieving higher ABR beyond that of BPSK input by the theoretical analysis and the simulations as well.  

\subsection{Theoretical Analysis} 
In the proposed method, the parallel transmission of $\textbf{x}_1$ and $\textbf{x}_2$ are encoded independent of each other.  The signal separation at the receiver is equivalent to tunnel two parallel channels to transmit the two signals separately. Thus, the overall ABR should be a problem of the summation accordingly.    

As has  been known, the concept of ABR is built upon maximum transmission bit rate, of which the BER should be infinitely  small. In comparison with the conventional coded modulations, the inner BPSK signals $\textbf{x}_1$ requires  error free of $\hat{\beta}$ for approaching to the channel capacity BPSK.  Thanks for Shannon theory of the channel coding that promises the existence of the channel codes to minimize the error probability of $C^{(2)}$ to infinitely small.  Consequently, the error probability of $\hat{v}^{(2)}$ can also be infinitely small because it relates to the info-bit in linear manner, so does $\hat{\beta}$.    

Thus, we can analyze the ABR of the inner BPSK signal $\textbf{x}_1$ with assumption of $\hat{\beta}=\beta$ in \eqref{equ-recove}.  The demodulation equation can be resolved into the following two  
\begin{eqnarray}
\begin{array}{l}\label{equ-x112}
y_1= x_1 + n_I     \ \ \ \    for \ \ \  \hat{\beta}=0 \\
y_1= x_1 + n_Q        \ \ \ \   for \ \ \  \hat{\beta}=\pi/2
\end{array} 
\end{eqnarray}
in forms of two scalar equations, where $ x_1 \in \{-\sqrt{E_s}, \sqrt{E_s}\}$,  $n_I$ and $n_Q$ are the two noise components perpendicular to $y$- and $x$ axis, respectively.   

It can be found that \eqref{equ-x112} is essentially the same as that of conventional BPSK with the same symbol energy and the same noise term of $n_I$ or $n_Q$, which are statistically same with variance of $\sigma_I^2=\sigma_Q^2=\sigma_N^2/2$.  Thus, it is found that ABR pertaining to $x_1$ can be the same as the channel capacity of BPSK input.     

Consequently, we can find that the ABR from transmission of
$\textbf{x}_2$ brings an extra ABR to the proposed method for beyond the capacity of BPSK.  The gain can be estimated based on the squared Euclidean distance and the noise term with respect to that of the inner BPSK.   The squared distance of the out-layer signal is found at $2E_s$ and the noise term is $\sigma_N^2$, while those of the inner BPSK signal are $4E_s$ and $\sigma_N^2/2$.  For having same bit-energy to noise ratio,  the code rate with $\textbf{x}_2$ can estimated by       
\begin{eqnarray}
\begin{array}{l}\label{equ-x111}
R_1/R_2 > \frac{8E_s/\sigma_N^2}{2E_s/\sigma_N^2} = 4  
\end{array} 
\end{eqnarray}
or
\begin{eqnarray}
R_2 < R_1/4
\end{eqnarray}
Thus, the additional ABR from $\textbf{x}_2$ can be roughly at 1/4 of the $\textbf{x}_1$.  Considering that the BER performance of $\textbf{x}_2$ should be much better than that of $\textbf{x}_1$ for preventing its error effects in \eqref{equ-recove}, the code rate of $\textbf{x}_2$ can be fewer than 1/8.      

The practical design will be studied by the simulations next. 

\subsection{Simulation Result}
In this subsection, the system simulations are performed, exactly along the procedures of section II, for showing the spectral efficiency gain of the proposed method with following specifications.    

The targeted spectral efficiency of the simulation can be set at
\begin{eqnarray}
\begin{array}{l}\label{x2-1}
\eta = R_1 + R_2
\end{array} 
\end{eqnarray}
where $\eta$ is the spectral efficiency of the proposed method, $R_1$ is the code rate used to $\textbf{x}_1$ and $R_2$ to $\textbf{x}_2$. 

By an appropriate selection of $R_1$ and $R_2$, and minimization of the BER to $10^{-8}$, the spectral efficiency $\eta$ in \eqref{x2-1} is regarded as the ABR of this approach.  Aiming at the BPSK modulation, we use $R_1=1/2$ to the channel code of $\textbf{x}_1$ and moderate code rate $R_2$ for the use to $\textbf{x}_2$ as explained in appendix.  It is found $R_2$ is a gain factor of this approach.         

For examining the performance, we use the bit-energy ratio to noise as the measurement.  Since the simulations are taken with ratio of symbol energy to the noise term, i.e., $E_s/N_0$, we need to make the conversion of 
\begin{eqnarray}
\begin{array}{l}\label{x2-2}
E_b/N_0 = E_s/N_0 - 10log_{10}(\eta) 
\end{array} 
\end{eqnarray}
to bit-energy to noise ratio, where $E_b$ is the bit-energy and $N_0$ is the corresponding noise term.  It is noted that the gain factor $R_2$ can contribute to $E_b/N_0$ as shown in \eqref{x2-2}.    

Before performing the system simulation, we address the following the problem of the error effect from $\hat{\beta}$ to the demodulation of $\textbf{x}_1$.     

A key to minimize the error rate of $\hat{\beta}$ is to reduce the error probability of the info-bits of $C^{(2)}$. For a given SNR, using the channel code having smaller code rate to $\textbf{x}_2$ can be an effective method. By such, the error rate of $\hat{v}^{(2)}$ or $\hat{\beta}$ can be significantly minimized accordingly.  However, reducing $R_2$ can withdraw its ABR contributions from $\textbf{x}_2$ to the total spectral efficiency as shown in \eqref{x2-1}.  Thus, a trade-off must be made.  

By doing the extensive simulations with the reference of the analysis on Euclidean space in the last subsection, we use one LDPC of code rate $R_2=1/12$ to  $\textbf{x}_2$ and another LDPC of $R_1=1/2$ to $\textbf{x}_1$ in the simulations.  The relevant issues are provided in the appendix. 

To be specified, we use the Matlab 2019a to perform the simulations, where the LDPC ( DVB-S.2) of code rate $R_1=1/2$ with 64800 code-bits and the constructed LDPC of code rate $R_2=1/12$ are used to $\textbf{x}_1$ and $\textbf{x}_2$, respectively.  The belief-passing algorithms are used for 50 iterations.  The results of BER performance are plotted as shown in Fig. 3, where we can find that bit-energy to noise ratio of 0.27 dB can reduce the BER to $10^{-8}$.  The gap between the simulation result and the channel capacity of BPSK at 0.5 bits/s/Hz is found at 0.08dB.  In comparison with the conventional BPSK using the same LDPC, a significant gain of 0.67dB gain is found. 

\begin{figure}[htb]
	\centering
	\includegraphics[width=0.4\textwidth]{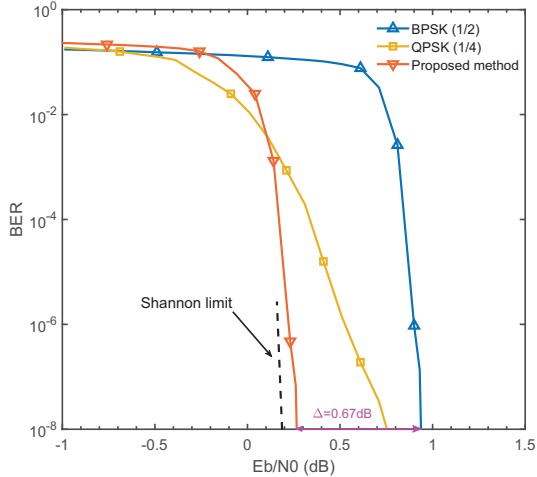}
	\caption{The BER comparison with BPSK and QPSK.}
	\label{Fig3}
\end{figure}

\begin{figure}[htb]
	\centering
	\includegraphics[width=0.4\textwidth]{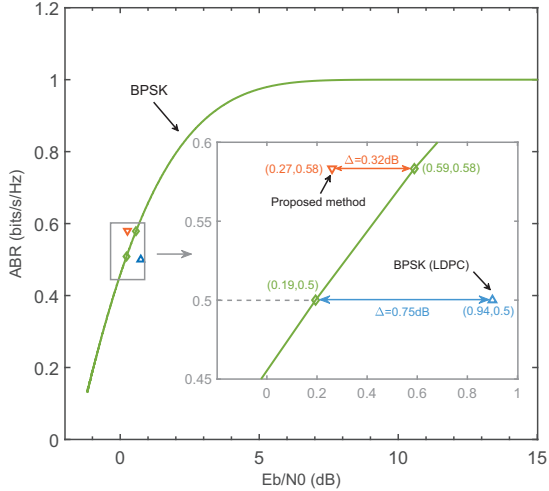}
	\caption{The spectral efficiency gain.}
	\label{Fig3}
\end{figure}

Moreover, considering the gain from $R_2$ in \eqref{x2-1}, i.e., at spectral efficiency of $1/2+1/12 \approx 0.58$ bits/s/Hz, we find 0.32 dB gain over the mutual information of BPSK input as shown in Fig. 4, where the vertical axis indicates the bits/s/Hz and horizontal axis the bit-energy to noise ratio.     

Besides, we provide the simulation results of BPSK modulation using LDPC DVB-S.2 in Fig. 4 as reference for showing the gain as well.  

Finally, in the recovery of $\hat{\beta}$, we use the results of $\hat{v}^{(2)}$ directly instead of using \eqref{equ-code-1} in all simulations of this paper for saving computing power of this research.

\section{Conclusion}

In this paper, we proposed a parallel transmission method, wherein the DMM is created for separating the parallel transmission signals for increasing the spectral efficiency. Both the theoretical analysis and the simulation results show the significant gain over the Shannon channel capacity of BPSK input.

\appendix
\renewcommand{\thefigure}{\Alph{figure}}
\section{Appendix}

This appendix explains the code construction of LPDC with code rate 1/12 and examines error effect of $\hat{\beta}$ to the transmission of $\textbf{x}_1$ at the BER performance. 

At first, we select LDPC (DVB-S.2) of code rate 1/3 and repeat each code-bit for 4 times to construct the code rate $R_2=1/12$.  

Secondly, we perform the simulation for showing the error effect of $\textbf{x}_2$ to demodulation of $\textbf{x}_1$ by 50 iterations.  The results are shown in Fig. A,  where the blue curve indicates the BER of $\textbf{x}_2$, the violet curve the BER unaffected by errors from $\textbf{x}_2$, i.e. error free of $\hat{v}^{(2)}$ or $\hat{\beta}$, and the yellow curve the BER affected by errors of $\textbf{x}_1$. we can found that the difference between the affected- and unaffected BER is insignificant at BER of $10^{-8}$ when code rate of $1/12$ is used to $\textbf{x}_2$. This means the transmission of signal $\textbf{x}_2$ does not affect signal $\textbf{x}_1$ for approaching is channel limit in the simulations.

\setcounter{figure}{0}
\begin{figure}[htb]
	\centering
	\includegraphics[width=0.4\textwidth]{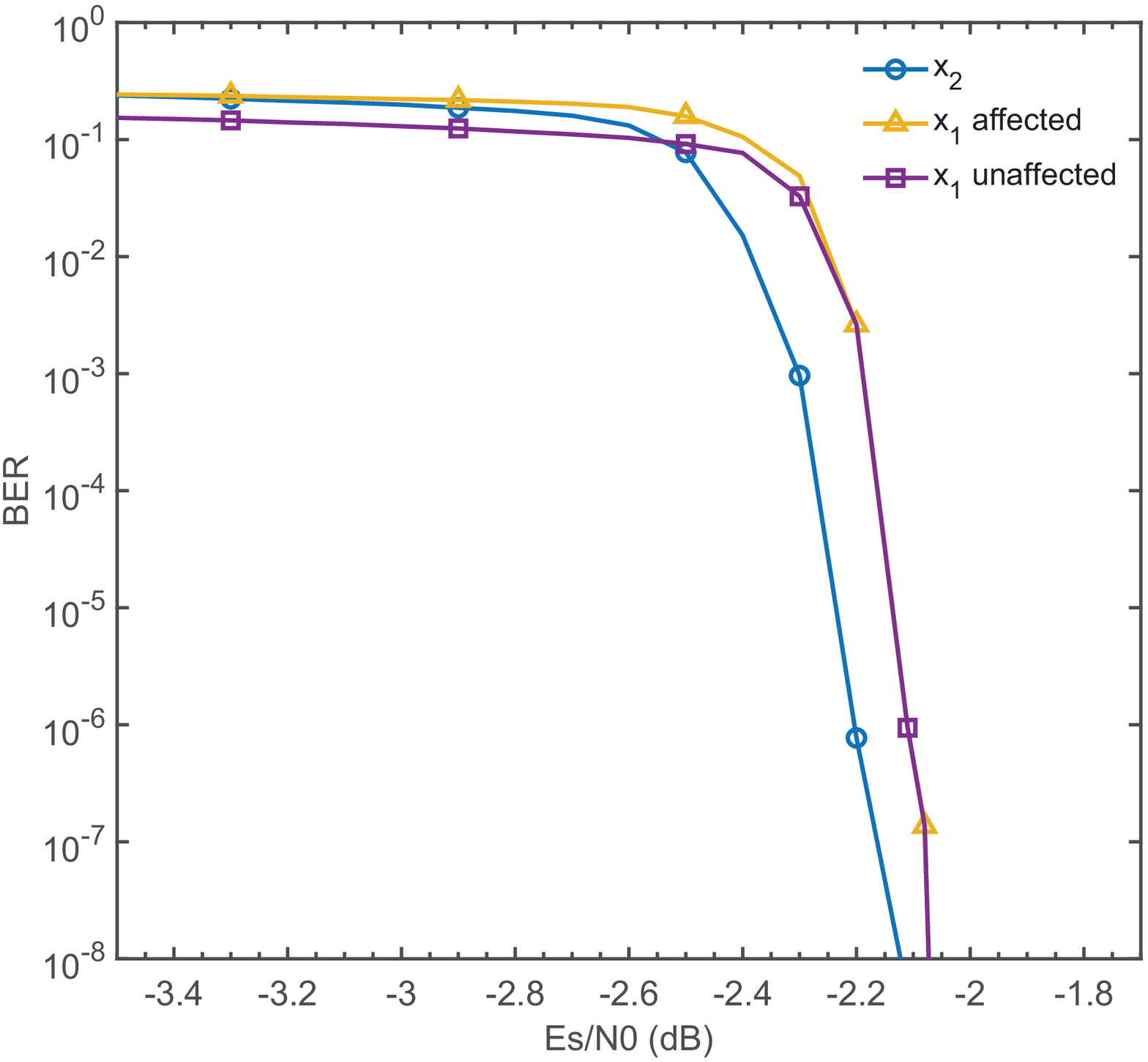}
	\caption{The BER performance with error effects from $\textbf{x}_2$.}
	\label{FIG_A}
\end{figure}

\end{document}